\documentclass[aps,prl,preprint,groupedaddress,longbibliography]{revtex4-1}
\usepackage{amsmath}
\usepackage{graphicx}
\usepackage{bm}
\allowdisplaybreaks

\begin{document}
\title{Kinetic Spin Decoherence in a Gravitational Field}
\author{Yue Dai}
\email[]{dy1983@gmail.com}
\affiliation{Beijing Computational Science Research Center, Beijing 100193, China}
\author{Yu Shi}
\email[]{yushi@fudan.edu.cn}
 \affiliation{Department of Physics  \&  State Key Laboratory of Surface Physics,  Fudan University, Shanghai 200433, China}
\affiliation{Collaborative Innovation Center of Advanced Microstructures, Fudan University, \\Shanghai 200433, China}


\begin{abstract}

We consider a  wave packet of a spin-1/2 particle  in a  gravitational field, the effect of which can be described in terms of a succession of local inertial frames. It is shown that integrating  out of the momentum yields a spin mixed state, with the entropy dependent on the deviation of metric from the flat spacetime. The decoherence occurs even if the particle is static in the gravitational field.
\end{abstract}


\maketitle


Spacetime tells matter how to move \cite{mtw}. For quantum objects, spacetime tells quantum states how to evolve, that is,  the quantum states are affected by the gravitational field. For example, by considering quantum fields near black holes, Hawking showed that black holes emit thermal particles \cite{hawking1974,hawking1975}. Analogously,  Unruh  showed that an accelerated detector in the Minkowski vacuum detects thermal radiation \cite{PhysRevD.7.2850,PhysRevD.14.870}.  In recent years, there have been a large number of discussions  about  quantum decoherence and quantum entanglement degradation in quantum fields~\cite{PhysRevD.68.085006, PhysRevLett.95.120404, PhysRevA.80.032315, PhysRevA.82.032324, PhysRevA.83.022314, PhysRevA.89.042336, PhysRevA.92.022334, Dai2015, PhysRevD.94.025012}. On the other hand, decoherence in the spin state due to the change of reference frame exists also for a single relativistic particle with spin degree of freedom. It was shown that  the spin entropy is not invariant under Lorentz transformation  unless the particle is in a momentum eigenstate~\cite{PhysRevLett.88.230402}. Such an  effect can be generalized to the presence of a gravitational field,  by considering a continuous succession of Lorentz transformations. This is a kinetic way of decoherence,  and occurs even if the particle is static.   It has been shown,   for a particle moving around a black hole near the Schwarzschild horizon,   that a  rapid spin decoherence is observed for an observer static in the Killing time~\cite{Ueda2005}. EPR correlation and the violation of Bell  inequality were also investigated \cite{PhysRevA.69.032113}. In this Letter, we  show that spin decoherence is quite general for a particle in a gravitational field even if the particle is static.

The state $\Psi_{p,\sigma}$ of a massive particle  is the eigenvector of the four-momentum $p^\mu$, and $\sigma$ denotes the spin. Under a homogeneous Lorentz transformation $\Lambda$ which carries the momentum from $p^\mu$ to $q^\mu = {\Lambda^\mu}_\nu p^\nu$, the state of the particle transforms as \cite{weinberg1995quantum}
\begin{equation}\label{statetransformation}
  U(\Lambda)\Psi_{p,\sigma} = \sqrt{\frac{(\Lambda p)^0}{p^0}} \sum_{\sigma '}D^{(1/2)}_{\sigma' \sigma}(W(\Lambda,p)) \Psi_{\Lambda p, \sigma'},
\end{equation}
where $D^{1/2}_{\sigma' \sigma}(W(\Lambda,p))$ is the spin-1/2 representation of  the Wigner rotaion $
  W(\Lambda,p)=L^{-1}(\Lambda p)\Lambda L(p)$, where
$L(p)$ represents a ``standard boost''  that carries the momentum from $(m,0,0,0)$ to $p^\mu$ and can be chosen as
\begin{eqnarray}
  {\left[L(p)\right]^\mu}_0 &=& {\left[L(p)\right]^0}_\mu = \frac{p^\mu}{m}, \label{standardboost1}\\
  {\left[L(p)\right]^i}_j &=& {\delta^i}_j + \frac{p^i p_j}{m(p^0+m)}, \label{standardboost2}
\end{eqnarray}
where $m$ is the mass of the particle, and indices $i,j=1,2,3$. It can be calculated that ${W^0}_0=1$, ${W^i}_0 = {W^0}_i =0$, and
\begin{equation}\label{wignerroation2}
{W^i}_j = {\Lambda^i}_j + \frac{{\Lambda^i}_0 p_j}{p^0 + m} - \frac{q^i {\Lambda^0}_j}{q^0 + m}.
\end{equation}

Now we calculate the Wigner rotation of a massive spin-1/2 particle moving in the gravitational field. Consider the spacetime metric
\begin{equation}\label{metric}
  ds^2=-(1+gx)^2dt^2+dx^2+dy^2+dz^2,
\end{equation}
which corresponds to the local coordinate system of an accelerated observer in flat space\-time or an observer in a uniform gravitational field \cite{mtw}. It can  also characterize the space\-time near the event horizon of a Schwarzschild black hole, as  the near horizon space\-time is locally flat for a large black hole, and static observers near the horizon in the Schwarzschild space\-time correspond to accelerated observers in Minkowski space\-time \cite{rindler2012essential,leonard2004introduction}.

We assume that the particle is moving with constant velocity $u^\mu = dx^\mu/d\tau$, that is, the spatial components of the velocity are independent of time. Note that the particle is not in free fall, it must be subject to a force, which may cause spin decoherence dynamically. We avoid this dynamical effect  by assuming  that the particle does not experience any torque. In order to calculate the Wigner rotation, we introduce local inertial frames (LIF) at each space\-time point along the particle's world line, by using a tetrad ${e_a}^\mu (x)$, defined through
\begin{equation}\label{tetrad1}
  {e_a}^\mu (x) {e_b}^\nu (x) g_{\mu\nu}(x)=\eta_{ab},
\end{equation}
where we use Greek letters to label the general-coordinates  $(t, x, y, z)$, and  Latin letters  to label  the coordinates of the LIF, with the components  denoted as $0, 1, 2, 3$.   For given $g_{\mu\nu} (x)$, the locally inertial coordinate at  $x_0$ can only be determined up to order $(x-x_0)^2$, as a LIF remains inertial after a Lorentz transformation \cite{weinberg1972gravitation}. But  the spin entropy is not invariant between different LIFs.

One choice of the LIF can be made by choosing
\begin{equation}
 {e_0}^t = \frac{1}{1+gx}, \quad {e_1}^x = {e_2}^y = {e_3}^z = 1,
\end{equation}
and all the other components are zero. In this LIF, the magnitudes of the spatial components of velocity of the particle are the same as those in the general coordinate, thus we may exclude the special relativistic effects described in \cite{PhysRevLett.88.230402}. The inverse of the tetrad can be obtained from
\begin{equation}
{e^a}_\mu(x){e_a}^\nu (x) = {\delta_\mu}^\nu, \quad {e^a}_\mu(x){e_b}^\mu (x) ={\delta^a}_b.
\end{equation}
And any vector $A^\mu$ is transformed  to
\begin{equation}
A^a = {e^a}_\mu A^\mu
\end{equation}
in the LIF. Now, if the particle does not move along a geodesic, the LIF updates  with time, so does the momentum of the particle $p^a$. The change of $p^a$ can be written as $\delta p^a = {\lambda^a}_b p^b d\eta$, where $\eta$ is the coordinate time of the LIF, and ${\lambda^a}_b$ is related to ${\Lambda^a}_b$ as  ${\Lambda^a}_b ={\lambda^a}_b d\eta + {\delta^a}_b$. ${\lambda^a}_b$ can be calculated to be
\begin{equation}\label{lambda}
 {\lambda^a}_b = - (a^a u_b - u^a a_b ) + {\chi^a}_b,
\end{equation}
where the first term is due to the acceleration of the particle, $a^a$ and $u^a$ are the acceleration and velocity in the LIF, respectively. $a^a = {e^a}_\mu a^\mu$, where
\begin{equation}
  a^\mu = u^\nu \nabla_\nu u^\mu = u^\nu \left[ \partial _\nu u^\mu + {\Gamma^\mu}_{\nu\sigma} u^\sigma \right],
\end{equation}
with  ${\Gamma^\mu}_{\nu\sigma}$ being  the Christoffel symbols, which  can be calculated according to the standard procedures, and $a_a$ can be obtained by lowering the indices. The results are
\begin{eqnarray}
  a^0 &=& - a_0 = \frac{gu^x\sqrt{1+u^2}}{1+gx}, \\
  a^1 &=& a_1 = \frac{g(1+u^2)}{1+gx},\\
  a^2 &=& a_2 = 0, \quad a^3 = a_3 =0.
\end{eqnarray}
The components of velocity in the LIF are
\begin{eqnarray}
  u^0 &=& - u_0 = \sqrt{1+u^2}, \\
  u^1 &=& u_1 = u_x, \quad u^2 = u_2 = u_y, \quad u^3 = u_3 = u_z.
\end{eqnarray}
The second term of Eq. (\ref{lambda}), ${\chi^a}_b$, which characterizes the change of the LIF along the world line, is given by
\begin{equation}
 {\chi^a}_b = u^\mu (x) \left[ {e_b}^\nu (x) \nabla_\mu {e^a}_\nu (x) \right].
\end{equation}
Thus ${\lambda^a}_b$ and ${\Lambda^a}_b$ can be  obtained.

Now we write ${W^i}_j$ as
\begin{equation}\label{wignerroation3}
  {W^i}_j = {\delta^i}_j + {\omega^i}_j,
\end{equation}
and then, by using Eq. (\ref{wignerroation2}) with $q^0 = p^0$ in our case, we obtain
\begin{eqnarray}
  {\omega^1}_2 &=& -{\omega^2}_1 = -\frac{g p_y \sqrt{m^2+p^2}}{(1+gx)m^2}, \\
  {\omega^1}_3 &=& -{\omega^3}_1 = -\frac{g p_z \sqrt{m^2+p^2}}{(1+gx)m^2}, \\
  {\omega^2}_3 &=& -{\omega^3}_2 = 0,
\end{eqnarray}
where $p = \sqrt{p^2_x + p^2_y + p^2_z}$. Note that $\omega$'s depend on $x$, which in turn depend on $p_x$. For a wave packet, $x$ may also has certain distribution, and we can simply replace $x$ as $\bar{x}$  and the initial   coordinate  $\bar{x}$   of the center of the wave packet, since it can be shown that if we take into account the width of the wave packet in $x$ direction, the difference is only of higher order. The spin-1/2 representation of the Winger rotation is
\begin{eqnarray}
  D^{1/2}_{\sigma'\sigma}(W(x)) &=& I + \frac{i}{2}\left[ \omega_{23}(x)\sigma_x + \omega_{31}(x)\sigma_y +\omega_{12}\sigma_z \right] d\eta \nonumber\\
  &=& I + \frac{i}{2}\left[{\bm{\omega}}(x)\cdot\bm{\sigma}\right]d\eta = I + \frac{i}{2}\omega (x) (\bm{\hat{n}}\cdot\bm{\sigma})d\eta,\label{infiniterep}
\end{eqnarray}
where $\omega(x)=\omega^2_{12}(x)+\omega^2_{23}(x)+\omega^2_{31}(x)$, $\hat{n}_i=\epsilon_{ijk}\omega_{jk}(x)/\omega(x)$, $\epsilon_{ijk}$ is the completely antisymmetric tensor with $\epsilon_{123}=1$, and $\bm{\hat{n}}=\hat{n}_i\bm{\hat{x}}^i$. For finite time $\eta$,
\begin{equation}
D^{1/2}_{\sigma'\sigma}(\eta) = T \left\{ \exp \left[ \frac{i}{2} \int_{0}^{\eta} \omega (x)(\hat{\bm{n}}\cdot \bm{\sigma}) d\eta\right] \right\},
\end{equation}
where $T$ represents time ordering. Hence
\begin{equation}
D^{1/2}_{\sigma'\sigma}(\eta) = I \cos\left(\frac{\theta}{2}\right) +i (\hat{\bm{n}}\cdot \bm{\sigma}) \sin\left(\frac{\theta}{2}\right),
\end{equation}
where
\begin{eqnarray}
  \theta &=& \int_{0}^{\eta} \omega(x) d\eta \nonumber\\
  &\approx& \frac{g\eta}{1+g\bar{x}} \frac{\sqrt{p_y^2 + p_z^2}\sqrt{m^2 + p^2}}{m^2} - \left( \frac{g\eta}{1+g\bar{x}}\right)^2 \frac{p_x \sqrt{p_y^2 + p_z^2}}{2m^2},
\end{eqnarray}
For a spin-1/2 particle, the initial state
$$
  \left( \begin{array}{c}
{a_0}( \bm{p} )\\
{b_0}( \bm{p} )
\end{array} \right).$$
becomes
\begin{equation}
  \left( \begin{array}{c}
{a}( \bm{p} )\\
{b}( \bm{p} )
\end{array} \right) = D^{1/2}_{\sigma'\sigma}(\eta) \left( \begin{array}{c}
{a_0}( \bm{p} )\\
{b_0}( \bm{p} )
\end{array} \right)\label{evolution}
\end{equation}
after the Wigner rotation.

Now we assume the initial wave packet is Gaussian,
\begin{eqnarray}
  a_0(\bm{p}) &=& \alpha \frac{\pi^{-3/4}}{\sqrt{w_x w_y w_z}} e^{-\frac{1}{2} \left[ \frac{\left( p_x - \bar{p}_x\right)^2}{w_x^2} + \frac{\left( p_y - \bar{p}_y\right)^2}{w_y^2} + \frac{\left( p_z - \bar{p}_z\right)^2}{w_z^2}\right]}, \\
  b_0(\bm{p}) &=& \beta \frac{\pi^{-3/4}}{\sqrt{w_x w_y w_z}} e^{-\frac{1}{2} \left[ \frac{\left( p_x - \bar{p}_x\right)^2}{w_x^2} + \frac{\left( p_y - \bar{p}_y\right)^2}{w_y^2} + \frac{\left( p_z - \bar{p}_z\right)^2}{w_z^2}\right]},
\end{eqnarray}
with $
 |\alpha|^2 + |\beta|^2 =1,$
$\bar{p}_i$  being  the average   momentum, and $w_i$ being  the width of the wave packet in each directions in the momentum space. Using Eq. (\ref{evolution}), we obtain that after rotation $\eta$, the state of the particle becomes
\begin{eqnarray}
  a(\bm{p}) &=& \frac{\pi^{-3/4}}{\sqrt{w_x w_y w_z}} e^{-\frac{1}{2} \left[ \frac{\left( p_x - \bar{p}_x\right)^2}{w_x^2} + \frac{\left( p_y - \bar{p}_y\right)^2}{w_y^2} + \frac{\left( p_z - \bar{p}_z\right)^2}{w_z^2}\right]} \nonumber\\
  &&\times\left[ \alpha \cos \left( \frac{\theta}{2}\right) + \beta \hat{n}_y \sin \left( \frac{\theta}{2} \right) + i \alpha \hat{n}_z  \sin \left( \frac{\theta}{2} \right)\right],\\
  b(\bm{p}) &=& \frac{\pi^{-3/4}}{\sqrt{w_x w_y w_z}} e^{-\frac{1}{2} \left[ \frac{\left( p_x - \bar{p}_x\right)^2}{w_x^2} + \frac{\left( p_y - \bar{p}_y\right)^2}{w_y^2} + \frac{\left( p_z - \bar{p}_z\right)^2}{w_z^2}\right]} \nonumber\\
  &&\times\left[ \beta \cos \left( \frac{\theta}{2}\right) - \alpha \hat{n}_y \sin \left( \frac{\theta}{2} \right) - i \beta \hat{n}_z  \sin \left( \frac{\theta}{2} \right)\right].
\end{eqnarray}
The density matrix of the particle can be written as
\begin{equation}
  \rho(\bm{p}_1,\bm{p}_2) = \left( {\begin{array}{*{20}{c}}
a(\bm{p}_1)a(\bm{p}_2)^*&a(\bm{p}_1)b(\bm{p}_2)^*\\
b(\bm{p}_1)a(\bm{p}_2)^*&b(\bm{p}_1)b(\bm{p}_2)^*
\end{array}} \right).
\end{equation}
The reduced density matrix of the spin can be obtained by tracing out over $\bm{p}$, that is, setting $\bm{p}_1 = \bm{p}_2 = \bm{p}$, and then integrating over $\bm{p}$. The result is
\begin{eqnarray}
  \int \left| a(\bm{p})\right|^2 d^3 p &=& |\alpha|^2 - \frac{\left( |\alpha|^2 - |\beta|^2\right)}{4 m^2} \left( \frac{g\eta}{1+g\bar{x}}\right)^2 \left( \bar{p}^2_z + \frac{w^2_z}{2}\right) \nonumber\\
  &&+ \frac{(\alpha \beta^*+\alpha^* \beta)}{2} \left[ \frac{g\eta}{1+g\bar{x}}\frac{\bar{p}_z}{m} - \left( \frac{g\eta}{1+g\bar{x}}\right)^2 \frac{\bar{p}_x\bar{p}_z}{m^2}\right]\nonumber\\
  &&- i \frac{(\alpha \beta^* - \alpha^* \beta)}{4m^2}\left( \frac{g\eta}{1+g\bar{x}}\right)^2\bar{p}_y\bar{p}_z, \label{densitymatrix1}\\
  \int \left| b(\bm{p})\right|^2 d^3 p &=& |\beta|^2 + \frac{\left( |\alpha|^2 - |\beta|^2\right)}{4 m^2} \left( \frac{g\eta}{1+g\bar{x}}\right)^2 \left( \bar{p}^2_z + \frac{w^2_z}{2}\right) \nonumber\\
  &&- \frac{(\alpha \beta^*+\alpha^* \beta)}{2} \left[ \frac{g\eta}{1+g\bar{x}}\frac{\bar{p}_z}{m} - \left( \frac{g\eta}{1+g\bar{x}}\right)^2 \frac{\bar{p}_x\bar{p}_z}{m^2}\right]\nonumber\\
  &&+ i \frac{(\alpha \beta^* - \alpha^* \beta)}{4m^2}\left( \frac{g\eta}{1+g\bar{x}}\right)^2\bar{p}_y\bar{p}_z,\\
  \int a(\bm{p}) b(\bm{p})^* d^3 p &=& \alpha\beta^* - \frac{(\alpha \beta^* + \alpha^* \beta)}{4m^2}\left( \frac{g\eta}{1+g\bar{x}}\right)^2 \left( \bar{p}^2_z + \frac{w^2_z}{2}\right) \nonumber\\
  &&- \frac{\alpha\beta^*}{2m^2} \left( \frac{g\eta}{1+g\bar{x}}\right)^2 \left( \bar{p}^2_y + \frac{w^2_y}{2}\right)\nonumber\\
  &&- \frac{\left( |\alpha|^2 - |\beta|^2\right)}{2} \left[ \frac{g\eta}{1+g\bar{x}} \frac{\bar{p}_z}{m} - \left( \frac{g\eta}{1+g\bar{x}}\right)^2\frac{\bar{p}_x \bar{p}_z}{m^2}\right]\nonumber\\
  &&- i\alpha\beta^* \left[ \frac{g\eta}{1+g\bar{x}} \frac{\bar{p}_y}{m} - \left( \frac{g\eta}{1+g\bar{x}}\right)^2\frac{\bar{p}_x \bar{p}_y}{m^2}\right]\nonumber\\
  &&+ i\frac{\left( |\alpha|^2 - |\beta|^2\right)}{4 m^2} \left( \frac{g\eta}{1+g\bar{x}}\right)^2 \bar{p}_y \bar{p}_z. \label{densitymatrix2}
\end{eqnarray}

Now we consider two cases. First we set $\alpha = 1$ and $\beta = 0$, the result is
\begin{eqnarray}
  \int \left| a(\bm{p})\right|^2 d^3 p &=& 1 - \frac{1}{4m^2} \left( \frac{g\eta}{1+g\bar{x}}\right)^2 \left( \bar{p}_z^2 + \frac{w_z^2}{2} \right), \\
  \int \left| b(\bm{p})\right|^2 d^3 p &=& \frac{1}{4m^2} \left( \frac{g\eta}{1+g\bar{x}}\right)^2 \left( \bar{p}_z^2 + \frac{w_z^2}{2} \right),\\
  \int a(\bm{p}) b(\bm{p})^* d^3 p &=& -\frac{1}{2m}\frac{g\eta}{1+g\bar{x}}\bar{p}_z + \frac{1}{4m^2} \left( \frac{g\eta}{1+g\bar{x}}\right)^2 \left( 2\bar{p}_x + i\bar{p}_y\right)\bar{p}_z.
\end{eqnarray}
The spin entropy can be calculated by using the definition
$
S = - \sum_{j} \lambda_j \ln \lambda_j,$
where $\lambda_j$'s are the eigenvalues of the density matrix. The result is
\begin{equation}\label{s1}
  S \approx \xi_z(1-\ln \xi_z),
\end{equation}
where $
  \xi_z = \frac{1}{2}\left[\frac{g w_z \eta}{2m(1+g\bar{x})}\right]^2.$
In Eq. (\ref{s1}), we have neglected the terms of higher orders in $\xi_z$. Eq. (\ref{s1}) is similar to  the results for Lorentz transformation~\cite{PhysRevLett.88.230402},
$$ S \approx \left( w^2 \tanh^2 \frac{\alpha}{2} /8m^2 \right) \left[ 1-\ln \left( w^2 \tanh^2 \frac{\alpha}{2} /8m^2 \right)\right],$$
where $\alpha = \cosh^{-1} (1-\beta^2)^{-1/2}$,  $w$ is the width of the momentum distribution.   The difference is that in our case, the spin entropy is generated by the gravity  and increases with time. It can be seen that our result does not depend on  the momentum $p^i$  of the particle. In other words, the spin entropy of the particle would increase with time even the particle is static. The widths of the wave packet in the momentum space appear in our result. In the limit of $\frac{w_z}{m} \to 0$, that is, if the particle is in a momentum eigenstate, $S$ remains zero.

We can also calculate the purity defined by $P = \rm{Tr}\,\rho^2$, the result is
\begin{equation}
P\approx 1 - 2\xi_z,
\end{equation}
which does not depend on the momentum either.

Now we consider the  case $\alpha=\beta=\frac{1}{\sqrt{2}}$. We obtain \begin{eqnarray}
  \int \left| a(\bm{p})\right|^2 d^3 p &=& \frac{1}{2} + \frac{1}{2m}\frac{g\eta}{1+g\bar{x}}\bar{p}_z - \frac{1}{2m^2} \left( \frac{g\eta}{1+g\bar{x}}\right)^2 \bar{p}_x \bar{p}_z, \\
  \int \left| b(\bm{p})\right|^2 d^3 p &=& \frac{1}{2} - \frac{1}{2m}\frac{g\eta}{1+g\bar{x}}\bar{p}_z + \frac{1}{2m^2} \left( \frac{g\eta}{1+g\bar{x}}\right)^2 \bar{p}_x \bar{p}_z, \\
  \int a(\bm{p}) b(\bm{p})^* d^3 p &=& \frac{1}{2} - \frac{1}{4m^2}\left( \frac{g\eta}{1+g\bar{x}}\right)^2 \left( \bar{p}_y^2 +\frac{w_y^2}{2}\right) - \frac{i}{2} \frac{g\eta}{1+g\bar{x}}\frac{\bar{p}_y}{m}\nonumber\\
  &&-\frac{1}{4m^2}\left( \frac{g\eta}{1+g\bar{x}}\right)^2 \left( \bar{p}_z^2 +\frac{w_z^2}{2}\right) + \frac{i}{2} \left( \frac{g\eta}{1+g\bar{x}}\right)^2 \frac{\bar{p}_x\bar{p}_y}{m^2}.
\end{eqnarray}
Thus the spin entropy is
\begin{equation}
  S \approx (\xi_y + \xi_z) (1-\ln (\xi_y + \xi_z)),
\end{equation}
where
\begin{equation}
  \xi_y = \frac{1}{2}\left[\frac{g w_y \eta}{2m(1+g\bar{x})}\right]^2.
\end{equation}
The purity  is calculated to be
\begin{equation}
  P \approx 1 - 2(\xi_y + \xi_z).
\end{equation}
Again, the spin entropy and the purity do not depend on the momentum, what determines the final results is the width of the momentum distribution.

Now we consider a static particle with the general values of  $\alpha$ and $\beta$. Using Eqs. (\ref{densitymatrix1})--(\ref{densitymatrix2}) with $p_i=0$, we obtain
\begin{eqnarray}
  \int \left| a(\bm{p})\right|^2 d^3 p &=& |\alpha|^2 - \left( |\alpha|^2 - |\beta|^2\right) \xi_z, \\
  \int \left| b(\bm{p})\right|^2 d^3 p &=& |\beta|^2 + \left( |\alpha|^2 - |\beta|^2\right) \xi_z,\\
  \int a(\bm{p}) b(\bm{p})^* d^3 p &=& \alpha\beta^* - 2\alpha\beta^* \xi_y- (\alpha \beta^* + \alpha^* \beta)\xi_z .
\end{eqnarray}
The spin entropy and the purity are, up to the first order of $\xi_y$ and $\xi_z$,
\begin{equation}
S \approx \xi \left( 1 - \ln \xi \right)
\end{equation}
and
\begin{equation}
P \approx 1 - 2\xi,
\end{equation}
respectively, where
\begin{equation}
\xi = 4 |\alpha|^2 |\beta|^2 \xi_y + \left(\alpha^2 +\beta^2\right)\left[(\alpha^*)^2+(\beta^*)^2\right] \xi_z.
\end{equation}

To summarize, we have calculated the spin entropy and the purity of  a spin-1/2 particle in the presence of the gravitational field. It has been shown that because of  the gravity, the spin entropy of the particle increases with time  even if the particle is static. This effect is generated by the gravity, and if $g\to 0$, the spin does not decohere. Our results also depend on the uncertainty of the momentum of the particle. If $\frac{w_i}{m}\to 0$, that is, the particle is in the momentum eigenstate, the spin entropy remains zero.

Although in our calculations a uniform gravitational field is employed, the results are quite general for any spacetime metrics. The calculations here are only related to the gravitational force felt by the particle, regardless the specific shape of the spacetime. For any particle moving in the general curved spacetime, we are able to calculate the Wigner rotation of the particle by introducing instantaneous LIFs, and hence we conclude that the spin state of the particle would decohere if the particle is not moving freely.

\begin{acknowledgments}
We thank Professor Chang-Pu Sun for helpful discussions. This work was supported by
National Basic Research Program of China (Grant No. 2016YFA0301201 \& No. 2014CB921403)
NSFC (Grant No. 11534002 \& No.  11574054)
NSAF (Grant No. U1730449 \& No. U1530401)
\end{acknowledgments}

\bibliography{spinh}

\end{document}